\documentclass[useAMS,usenatbib]{mn2e}
\usepackage{graphicx,lscape,amssymb}
\usepackage{rotating}
\usepackage{natbib}
\usepackage{epstopdf}
\usepackage{fancyhdr}

\newcommand{\Teff}{\mbox{$T_{\mathrm{eff}}$}}
\newcommand{\Msun}{\mbox{$\mathrm{M}_{\odot}$}}
\newcommand{\Nwd}{245}
\newcommand{\newwd}{188}

\title[SDSS white dwarf candidates in LAMOST DR3]{An independent test of the photometric selection of white dwarf candidates using LAMOST DR3}

\author[Gentile Fusillo et al.]{N. P. Gentile Fusillo$^1$, A. Rebassa-Mansergas$^2$,
  B.T. G\"ansicke$^1$,  X.-W.
  Liu$^{2,3}$, \newauthor J.J. Ren$^{3}$, D. Koester$^{4}$,Y. Zhan$^{5}$, Y. Hou$^{5}$, Y. Wang$^{5}$, M. Yang$^{6}$ \\
$^1$ Department of Physics, University of Warwick, Coventry, CV4 7AL,
  UK\\
$^2$ Kavli Institute for Astronomy and Astrophysics, Peking
  University, 100871 Beijing, P.\,R.\,China\\
$^{3}$ Department of Astronomy, Peking University, Beijing 100871,
P.\,R.\,China\\
$^{4}$ Institut f\"ur Theoretische Physik und Astrophysik, University
  of Kiel, 24098 Kiel, Germany\\
$^{5}$ Nanjing Institute of Astronomical Optics \& Technology, National Astronomical Observatories,\\ \- \- Chinese Academy of Sciences, Nanjing 210042, P.\,R.\,China\\
$^{6}$ Key Laboratory of Optical Astronomy, National Astronomical Observatories,\\
\- \- Chinese Academy of Sciences, Beijing 100012, P.\,R.\,China}

\begin{document}
\maketitle

\label{firstpage}

\begin{abstract}
In \citet{gentilefusilloetal15-1} we developed a selection method for white dwarf candidates which makes use of photometry, colours and proper motions to calculate a \emph{probability of being a white dwarf} ($P_\mathrm{WD}$). The application of our method to the Sloan Digital Sky Survey (SDSS) data release 10  resulted in $\simeq 66,000$ photometrically selected objects with a derived $P_\mathrm{WD}$, approximately $\simeq21000$ of which are high confidence white dwarf candidates.
Here we present an independent test of our selection method based on a sample of spectroscopically confirmed white dwarfs from the LAMOST (Large Sky Area Multi-Fiber Spectroscopic Telescope) survey. We do this by cross matching all our $\simeq 66,000$ SDSS photometric white dwarf candidates with the over 4 million spectra available in the third data release of LAMOST. This results in 1673 white dwarf candidates with no previous SDSS spectroscopy, but with available LAMOST spectra. Among these objects we identify 309 genuine white dwarfs. We find that our $P_\mathrm{WD}$ can efficiently discriminate between confirmed LAMOST white dwarfs and contaminants. Our white dwarf candidate selection method can be applied to any multi-band photometric survey and in this work we conclusively confirm its reliability in selecting white dwarfs without recourse to spectroscopy.  We also discuss the spectroscopic completeness of white dwarfs in LAMOST, as well as deriving effective temperatures, surface gravities and masses for the hydrogen-rich atmosphere white dwarfs in the newly identified LAMOST sample. 

\end{abstract}

\begin{keywords}
(stars:) white dwarfs - surveys - catalogues
\end{keywords}

\section{Introduction}

Main sequence stars of masses $M\lesssim8-10.5 M_\odot$ are destined
to become white dwarfs (\citealt{ibenetal97-1}, \citealt{smarttetal09-1}). White dwarfs are
therefore the most common stellar remnants in the Galaxy.  However, because of their small radii, and hence low luminosities,
constructing a large, homogeneous and unbiased sample of white dwarfs
is still an ongoing challenge.

The advent of modern large scale observational surveys has allowed to
constrain fundamental parameters such as the white dwarf space density
\citep{limoges+bergeron10-1, sionetal14-1}, mass distribution and mass
function \citep{bergeronetal92-1, liebertetal05-1, kepleretal07-1,
  falconetal10-1, tremblayetal13-1, Kleinmanetal13-1}, luminosity
function \citep{oswaltetal96-1, degennaroetal08-1, torresetal14-1,
  cojocaruetal14-1} and formation rate \citep{huetal07-1,  verbeeketal13-1}.  However, the white dwarf samples used in these studies are
inevitably affected by selection effects, and it is difficult to quantify how and to what extent the observational biases affect the derived results. 
We recently developed a selection method which enables us to identify high-confidence white dwarf candidates in large multi-colour photometric surveys. 
The application of our method to
SDSS data release (DR) 10 allowed us to investigate the spectroscopic biases of SDSS DR10 and  resulted in a catalogue of 65,768 bright
(g\,$\leq19$) point sources  \citep{gentilefusilloetal15-1}  with computed \emph{probability of being a white dwarf} ($P_\mathrm{WD}$). Using our catalogue it is possible to select
$\sim$14,000 high-confidence white dwarf candidates that have not
yet received any spectroscopic follow up, which is ultimately needed to measure their fundamental parameters such as mass and cooling age. Spectroscopic follow-up
will also help in identifying rare white dwarf types, which are key
objects for a wide variety of studies such as exploring the late
evolutionary stages of a wide range of progenitor stars
\citep{schmidtetal99-1, dufouretal10-1, gaensickeetal10-1}, searching
for low-mass companions \citep{farihietal05-1, girvenetal11-1,
  steeleetal13-1}, metal polluted white dwarfs \citep{sionetal90-1,
  zuckermanetal98-1, dufouretal07-2, koesteretal14-1} and white dwarfs
with dusty or gaseous planetary debris discs \citep{gaensickeetal06-3,
  farihietal09-1, debesetal11-2, wilsonetal14-1}.

\begin{figure}
\includegraphics[width=\columnwidth]{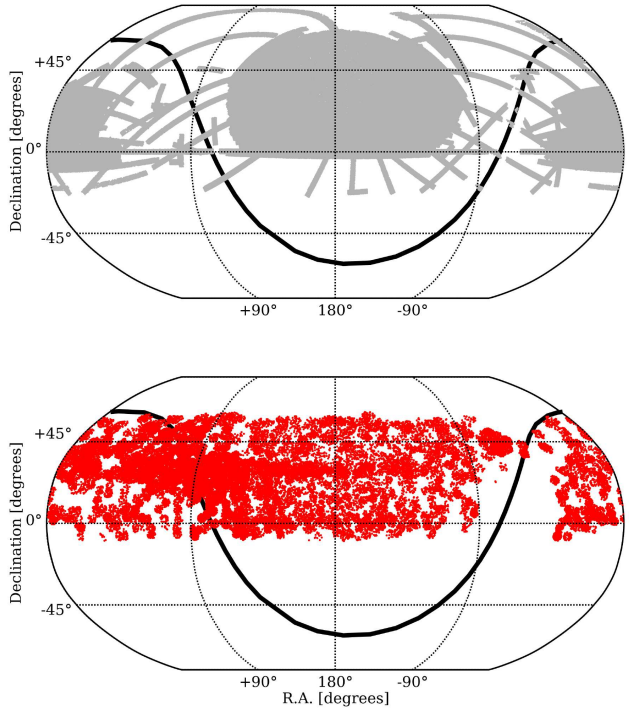}
\caption{\label{skycover}Photometric coverage of SDSS DR10 (top panel)
  and LAMOST DR3 pointings (bottom panel). The black line indicates
  the location of the Galactic plane.}
\end{figure}

The recently initiated Large Sky Area Multi-Object Fiber Spectroscopic
Telescope (LAMOST; \citealt{luoetal15-1}) survey provides hundreds of thousand of spectra per
year and its sky coverage overlaps to a large degree with the
photometric footprint of SDSS (Fig\,\ref{skycover}). In this work
we crossmatch our catalogue of photometric SDSS white dwarf candidates
with the over 4 million spectra obtained to date by LAMOST and find
spectroscopy spectra for 309 white dwarfs. We use this new set of
white dwarfs to test our photometric selection method, as well as to
analyse the effects of the target selection algorithm of LAMOST on the
observed white dwarf population. We also provide the stellar
parameters of the newly-identified DA (hydrogen dominated atmosphere)
white dwarfs, namely the effective temperatures, surface gravities and
masses.

\section{The LAMOST survey}
\label{lamost}

LAMOST is a quasi-meridian reflecting Schmidt telescope of effective
aperture $\sim$4m located at Xinglong Observing Station in the Hebei
province of China \citep{cuietal12-1}.  LAMOST uses 16 fiber-fed
spectrographs each equipped with a red and a blue channel CDD
camera. Each spectrograph counts with 250 fibers, thus LAMOST is able to
obtain a total of 4,000 simultaneous spectra. The wavelength coverage
of the spectra is $\sim$3800-9000\,\AA \- at a resolving power of
$\sim$1,800.  Although the flux calibration of the LAMOST spectra is
relative \citep{songetal12-1}, the spectral energy distribution is
correctly characterized and classifications based on visual inspection
of the spectra can be considered as reliable.

LAMOST started operation in 2009 and began a five-year regular survey
in 2012.  This survey consists of two main parts with different
science goals and target selection criteria \citep{zhaoetal12-1}.  The
LAMOST Extra-Galactic Survey (LEGAS) studies the large scale structure
of the Universe. The LAMOST Experiment for Galactic Understanding and
Exploration (LEGUE) focuses on characterizing the structure and
evolution of the Milky Way \citep{dengetal12-1} and is sub-divided
into three sub-surveys \citep{carlinetal12-1, chenetal12-1,
  liuetal14-1, yuanetal15-1}: the spheroid, the disk, and the Galactic
anti-center.

The current number of available LAMOST spectra is $\sim$4.6
million. These include the full second data release (DR2) plus the first three months of data of the third data release (DR3).

\section{SDSS photometric white dwarf candidates observed by LAMOST}

The catalogue of \citet{gentilefusilloetal15-1} includes 65,768 bright
(g\,$\le$19) point sources selected according to their available SDSS
photometry and colours. The catalogue contains not only white dwarfs,
but also quasars and other blue stars, however each object has an
associated $P_\mathrm
{WD}$ calculated from its $g-z$ colour and reduced proper
motion.


We estimate that our catalogue contains $\sim$14,000 white dwarfs
which have not yet received spectroscopic follow up in the latest data
release of SDSS (DR12).  The vast number of available LAMOST spectra,
combined with the large overlap in the SDSS and LAMOST footprints
(Fig.\,\ref{skycover}), implies that a considerable number of white dwarf
candidates have most likely been observed by LAMOST.  Consequently
combining our catalogue of SDSS white dwarfs candidates with all
available LAMOST spectra is not only a quick and reliable way to
identify new white dwarfs, but also provides a further test to
corroborate the reliability of our selection method.

We cross-matched all 65,768 objects from	 the
\citet{gentilefusilloetal15-1} catalogue with the list of the 4.6 million LAMOST
spectra and retrieved 6,101 spectra corresponding to 5173
unique objects (Table\,\ref{recap}).  3500 of these have also received SDSS spectroscopic follow
up and 64 further objects had already been identified on the base of their LAMOST spectra as white dwarfs or white
dwarf binaries by \citet{zhangetal13-1, zhaoetal13-1, renetal14-1,
  rebassa-mansergasetal15-1}.

\begin{table}
\caption{\label{recap} Summary of the cross-matching of SDSS white dwarf candidates with LAMOST DR3}
\begin{center}
\begin{tabular}{rcc}
\hline
 & n. of objects & n. of spectra\\
\hline
All objects from cross-match & $5173$ & $6101$ \\
& & \\
SDSS and LAMOST spectra & $3500$ & $3964$ \\
of which WDs & $774$ & $1177$ \\
& & \\
LAMOST spectra only & $1673$ & $2137$ \\
of which WDs & $309$ & $387$\\
already published & $64$ & $97$ \\
unpublished & $245$ & $290$ \\
\hline
\end{tabular}
\end{center}
\end{table}

\begin{figure*}

\includegraphics[width=\textwidth]{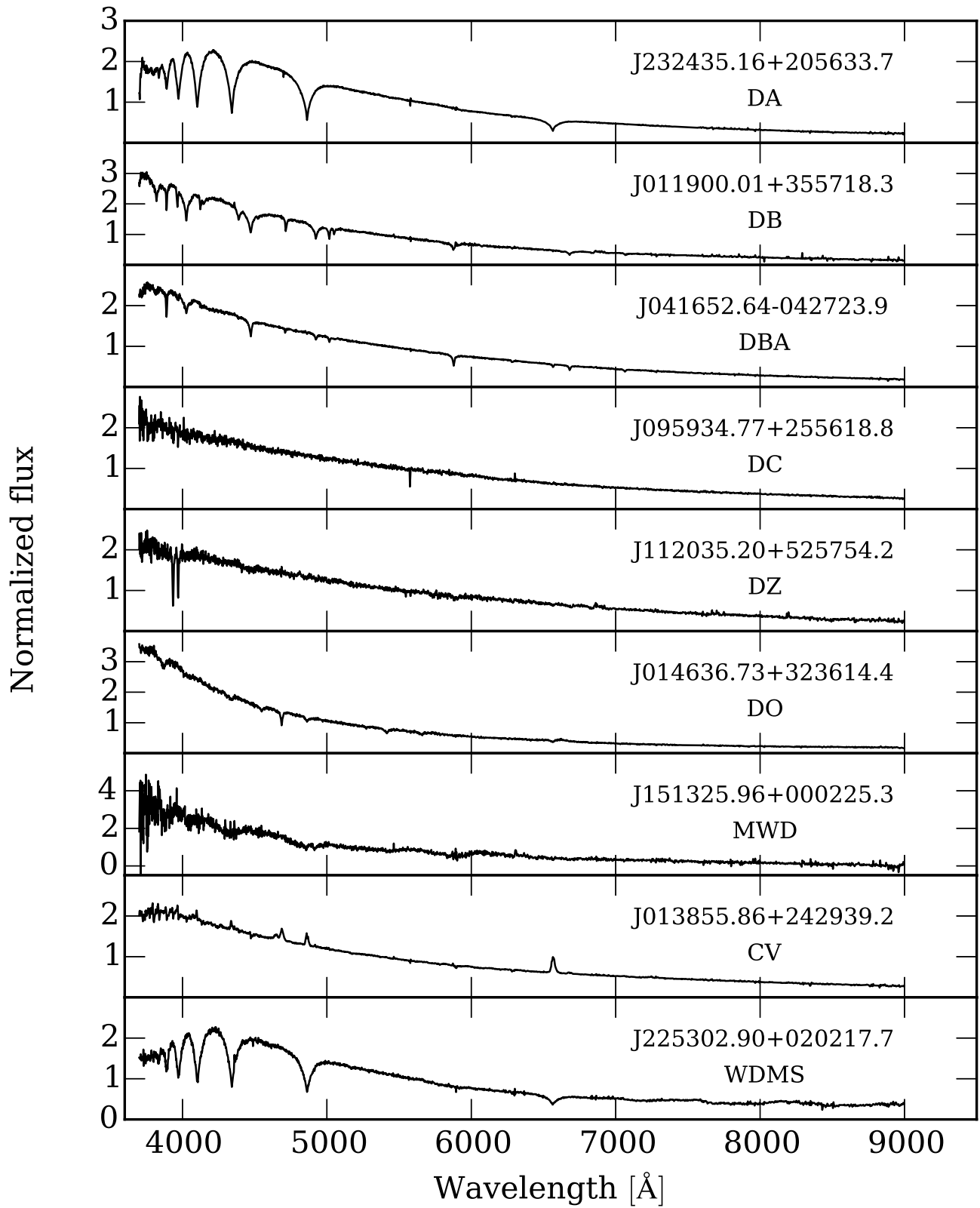}
\caption{\label{sample_spec} Sample LAMOST spectra of different types of white dwarfs}
\end{figure*}

Since the main goal of this project is to test
our photometric selection method using an independent spectroscopic
sample of white dwarfs, we limited ourselves to objects  which have no SDSS spectroscopic counterpart.  This reduced
the sample to 2,040 spectra corresponding to 1,609 unique LAMOST objects; plus the 64 known LAMOST white dwarfs mentioned above. \textbf{Finally we visually classified all remaining white dwarf candidates with unpublished LAMOST spectra}.  We subdivided the identified white
dwarfs into 10 types (Table \ref{class}, Fig.\,\ref{sample_spec}), and the contaminants into 3
types: ``K/M stars", ``Quasars" (QSOs) and ``Narrow Line
Hydrogen Stars" (NLHS), in which we group different stars with
low-gravity hydrogen dominated atmospheres such as subdwarfs, extreme
horizontal branch stars and A/B type stars. We also marked as
``unreliable" all spectra which had a signal-to-noise ratio too low for a reliable classification.  The results of our
classification are summarized in Table\,\ref{class}.

\begin{table}
\caption{\label{class} Classification of the 2,040 unpublished LAMOST
  spectra of 1,609 white dwarf candidates from the
  \citet{gentilefusilloetal15-1} catalogue.}
\begin{center}
\begin{tabular}{lcc}
\hline
Class & n. of objects & n. of spectra\\
\hline
DA & $196$ & $222$ \\
DB & $5$ & $6$ \\
DAB/DBA & $2$ & $2$ \\
DO & $3$ & $5$ \\
DC & $10$ & $12$ \\
DZ & $3$ & $3$\\
Magnetic WD & $2$ & $2$ \\
WD+MS & $4$ & $4$ \\
CV & $19$ & $33$ \\
Planetary nebula & $1$ & $1$ \\
NLHS & $393$ & $538$\\
QSO & $546$ & $678$\\
K/M stars & $3$ & $3$ \\
Unreliable & $422$ & $531$ \\

\hline
\end{tabular}
\end{center}
\end{table}

Inspection of the Table reveals that we have identified \Nwd\,
white dwarfs (with a total of 290 spectra), of which over 80 per cent
have hydrogen dominated atmospheres (DA).  DAs are known to constitute
the vast majority of all white dwarfs \citep{mccook+sion99-1}, so the ratio above is unsurprising. 
Querying the SIMBAD astronomical database, we find that 57 of our \Nwd\, white dwarfs had already been identified as such (as single stars or part of binary systems) in other studies. In conclusion we report the discovery of \newwd\- new white dwarfs.

\subsection{Cataclysmic variables in the LAMOST sample}
During the classification of LAMOST spectra we identified 19 cataclysmic variables (CV, Table\,\ref{class}). 10 of these objects are known CVs  while the remaining nine are new discoveries. Considering the limited size of our white dwarf sample the number of CVs found is remarkably high. However, we are not aware of any aspects of the targeting strategy of LAMOST which could have led to preferential observation of CVs. A possible cause of this bias towards CVs may be that the strong emission features which characterize CVs can be easily recognized even in very noisy spectra. Indeed we find that most of the inspected LAMOST spectra of CVs are of low quality. 
In order to verify our classification of these objects we cross-matched our nine new CVs with the second data release of the Catalina survey (CSDR2, \citealt{drakeetal09-1}). CSDR2 provides multi-epoch photometry for over 500 million objects and has been an extremely useful resource for various campaign which searched for CVs (e.g. \citealt{breedtetal14-1}, \citealt{drakeetal14-1}). 
Two of our new CVs are not within the CSDR2 sky footprint, but we were able to recover and inspected Catalina light curves for the remaining seven (Table\,\ref{crts}). 
A complete analysis of these light curves is beyond the scope of this article, but we conclude these objects indeed show variability compatible with that of a CV.
\begin{table}
\caption{\label{crts} LAMOST objects identified as previously unknown CVs.}
\begin{center}
\begin{tabular}{lc}
\hline
Name & CRDR2\\
 & lightcurve\\
\hline
J003005.80+261726.3 & yes\\
J010903.02+275010.0 & yes\\
J013317.01+305329.8 & yes\\
J013855.86+242939.2 & yes\\
J052602.79+285121.3 & no\\
J062402.64+270410.2 & no\\
J074037.68+254109.4 & yes\\
J171630.84+444124.5 & yes\\
J172308.28+392455.2 & yes\\
\hline
\end{tabular}
\end{center}
\end{table}

\section{An independent test of the  Gentile Fusillo et al. 2015 white dwarf selection method}

\begin{figure}
\includegraphics[width=\columnwidth]{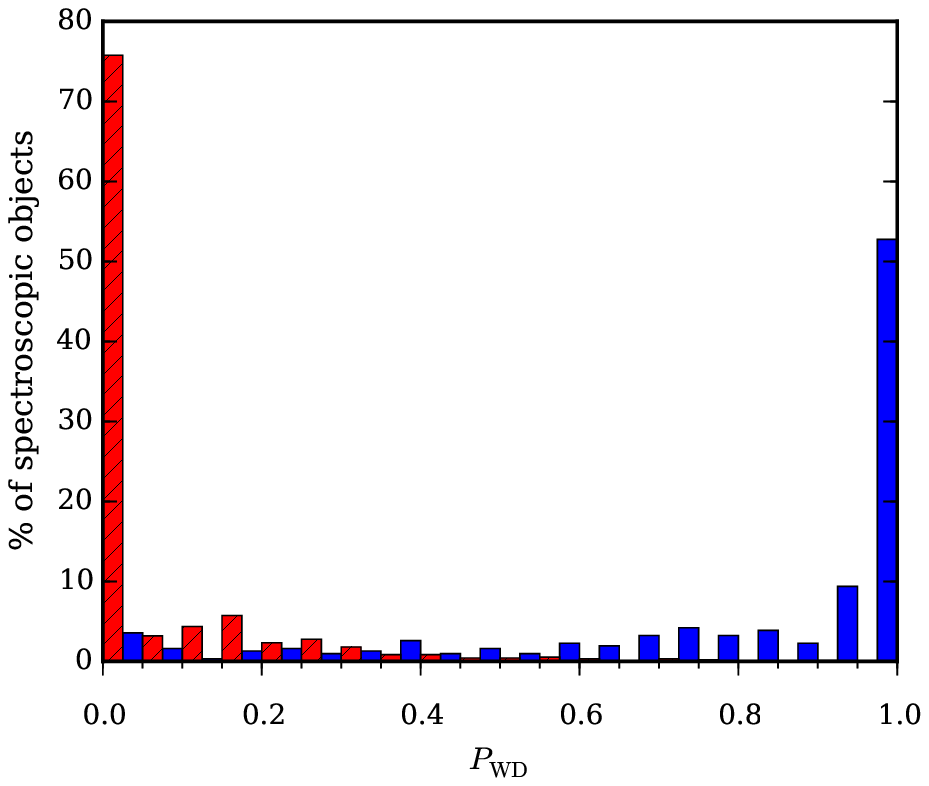}
\caption{\label{histo} Distribution of all spectroscopically confirmed
  LAMOST white dwarfs (blue) and contaminants (red, shaded) as a
  function of $P_\mathrm {WD}$.}
\end{figure}

In \citet{gentilefusilloetal15-1} we tested the reliability of our
white dwarf candidate selection method using a sample of 6,706
spectroscopically confirmed white dwarfs and over 20,000 contaminants
with available SDSS spectra.  For a given $P_\mathrm {WD}$ threshold,
we defined \textit{completeness} as the ratio of the number of white
dwarfs in the spectroscopic sample with at least that associated
probability to the total number of white dwarfs in the sample.
Similarly \textit{efficiency} was defined as the ratio of the number
of white dwarfs selected by the probability cut to the number of all
the objects retrieved by such selection. We concluded that our $P_\mathrm
{WD}$ can be reliably used to discern between white dwarfs and
contaminants. For example, selecting all objects with a $P_\mathrm
{WD} \geq 0.41$ resulted in a sample of white dwarf candidates which
is 95 per cent complete and 89.7 per cent efficient.
One of the main strengths of the selection method of \citet{gentilefusilloetal15-1} is that, even though it was developed using SDSS, it can in principle be applied to any large area survey which provides multi band photometry (e.g. VST ATLAS, APASS, SkyMapper, Pan-Starrs). The possibility of applying the selection method to surveys other than SDSS, however, stresses the need to test the robustness of our $P_\mathrm {WD}$ values on a sample of spectroscopically confirmed white dwarfs completely independent from SDSS.
Even though our sample of 309 (64 known ones and \Nwd\- identified as part of this work) confirmed LAMOST white dwarfs and 876
contaminants is small compared to the SDSS spectroscopic sample, it
provides a welcome opportunity to verify the reliability of our
$P_\mathrm {WD}$ values.  Figure\,\ref{histo} clearly shows that over
$\sim$80 per cent of all LAMOST contaminants have $P_\mathrm {WD} <
0.2$ and virtually no contaminant has $P_\mathrm {WD} >
0.5$. Similarly, the vast majority of LAMOST white dwarfs have
$P_\mathrm {WD} > 0.5$. Further inspection of Figure \ref{histo}
reveals also that $\sim$10 per cent of the LAMOST white dwarfs have
$P_\mathrm {WD} <0.4$ and would be missed by the most reasonable
selections based on $P_\mathrm {WD}$, i.e. a probability cut at $P_\mathrm {WD} \geq 0.41$. 
Inspection of these low probability objects reveals that the vast majority of these are CVs (see Sect 3.1). CVs have peculiar colours distinct from those of most single white dwarfs and the selection method of \citet{gentilefusilloetal15-1} is not optimized for them. 
Nonetheless  the statistic we compute on our LAMOST white dwarfs sample confirms the reliability of our selection method. The $P_\mathrm {WD}$ can confidently be used to select different samples white dwarfs according to the specific work one intends to carry out.



\section{Spectroscopic completeness of the white dwarfs identified by LAMOST}

\begin{figure*}
\includegraphics[width=2\columnwidth]{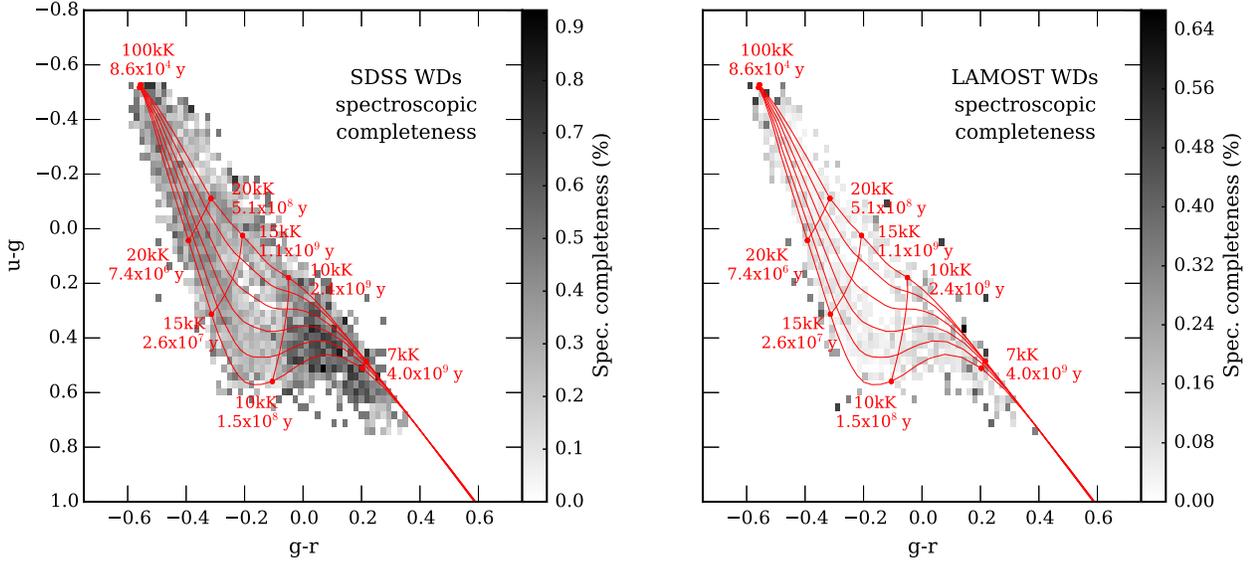}
\caption{\label{colour_cover} Spectroscopic completeness of SDSS white
  dwarfs (left panel) and LAMOST white dwarfs over the SDSS footprint (right panel). The values were computed as the ratio of spectroscopically
  confirmed white dwarfs to all high-confidence white dwarf candidates
  ($P_\mathrm {WD}$ $\geq 0.41$) within the $(u-g,g-r)$ colour-colour
  selection used in \citet{gentilefusilloetal15-1}.
  To correctly compute the spectroscopic completeness of LAMOST over the SDSS footprint, we used the entire sample of 1083 white dwarfs resulting from our initial cross match of LAMOST targets with the catalogue of \citet{gentilefusilloetal15-1} (newly identified LAMOST WDs + already published LAMOST WDs + LAMOST WDs with SDSS spectra). 
   White dwarf
  cooling tracks from \citet{holberg+bergeron06-1} are shown as
  overlay (red solid lines). The left panel clearly shows an area of
  higher spectroscopic completeness caused by the SDSS target
  selection algorithm, which favours the observations of QSOs (see
  sect.5).}

\end{figure*}

\begin{figure}
\includegraphics[width=\columnwidth]{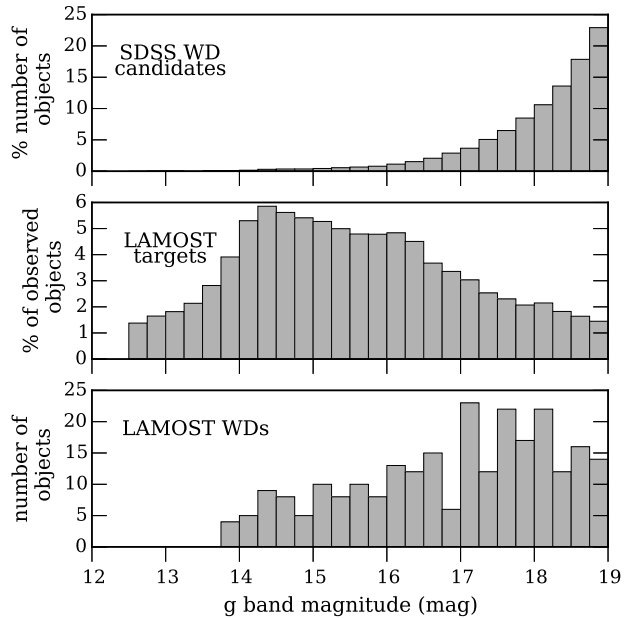}
\caption{\label{g_dist}\emph{Top panel}: $g$-band magnitude distribution of all SDSS white dwarf candidates from \citet{gentilefusilloetal15-1}.
\newline \emph{Middle panel}: $g$-band magnitude distribution of a random sub-set of 100,000 objects with LAMOST spectroscopy.
\newline \emph{Bottom panel}: $g$-band magnitude distribution of the newly identified LAMOST white dwarfs.}
\end{figure}

In \citet{gentilefusilloetal15-1} we find that, to date, the SDSS
white dwarf spectroscopic sample is $\sim$40 per cent complete.
However, this number is averaged over the entire SDSS photometric
footprint, large areas of which have not yet received any
spectroscopic follow up.  Furthermore, as we show in
Fig. \ref{colour_cover} (left panel), the spectroscopic completeness
of SDSS white dwarfs is also very colour-dependent. For example,
the spectroscopic completeness is highest for relatively cool white dwarfs with colours
similar to those of quasars, a simple consequence of the target
selection algorithm of SDSS. The LAMOST observing strategy differs from
SDSS's both in terms of area of sky covered and target selection.
Even though white dwarfs are only serendipitous or secondary targets
for both surveys, the resulting spectroscopic samples are, to a
certain degree, complementary.  Figure \ref{colour_cover} (right
panel) illustrates that the spectroscopic completeness of LAMOST over the SDSS footprint does
not dramatically depend on colour, although it does rise slightly in quasar-dominated areas.   
Naturally the number of white dwarfs increases at higher magnitudes as a larger volume is observed. However LAMOST observed mostly bright objects ($g\simeq 14-16$, Fig.\,\ref{g_dist}) and therefore did not target the vast majority of photometrically selected white dwarfs.
Consequently, even though the LAMOST white dwarf sample is less biased by colour it is limited in size, and the overall white dwarfs spectroscopic completeness is much lower than the SDSS one.

In order to compute the white dwarfs spectroscopic completeness of LAMOST we used the entire sample of 1083 white dwarfs resulting from our initial cross match of LAMOST targets with the catalogue of \citet{gentilefusilloetal15-1} (newly identified LAMOST WDs + already published LAMOST WDs + LAMOST WDs with SDSS spectra). 
However it is important to keep in mind that the final sample of new white dwarfs presented is not representative of all the white dwarfs observed by LAMOST. The catalogue of white dwarf candidates of \citet{gentilefusilloetal15-1} can only be considered complete white dwarfs \textbf{with available proper motions,  $T_\mathrm {eff} \gtrsim 8000$ K  and $g \leq19$}. Consequently the LAMOST white dwarf sample discussed here is affected by these same limitations.
Even though cool white dwarf are particularly faint and LAMOST mostly targeted bright objects ($g\simeq 14-16$), we expect that among the 4.6 million spectra collected to date there should be some cooler nearby white dwarfs which were not included in this work.
Finally, LAMOST has also extensively covered areas of the sky which lie outside the SDSS footprint (e.g. the Galactic anti-center;\citealt{yuanetal15-1}) and any white dwarf observed in those areas would, by definition, not be included in the catalogue presented here.

\section{Stellar parameters}

In this section we derive the stellar parameters of the newly
identified LAMOST DA white dwarfs (Table \ref{class}). We do this
following the fitting routine developed by
\citet{rebassa-mansergasetal07-1, rebassa-mansergasetal10-1} adapted
to LAMOST spectra \citep{rebassa-mansergasetal15-1}. A brief
description of this procedure is given here, and we refer the reader
to the references above for further details.

To determine the effective temperature ($\Teff$) and surface gravity ($\log g\,\mathrm{[g	 \, cm \, s^{\textrm{-}2}]}$) we fit the normalised H$\beta$ to H$\epsilon$ line profiles of
each spectrum with the DA model grid of \citet{koester10-1} using a mixing-length parameter (ML2/$\alpha$) of 0.6. 
Since the equivalent widths of the Balmer lines go through
a maximum near $\Teff=13,000$\,K, the line fitting provides two
possible solutions, i.e.``hot'' and ``cold'' solutions. We break this
degeneracy fitting the entire white dwarf spectrum (continuum plus
lines). The continuum spectrum of a DA WD is mostly sensitive to
$\Teff$, therefore the best-fit value from the entire spectrum
generally indicates which of the two solutions is the most reliable
one.  However, because of uncertainties in the LAMOST flux calibration
(Section \ref{lamost}), the best-fit to the entire spectrum may be
subject to systematic uncertainties.  Thus, the choice between hot and
cold solution is further guided by comparing the ultraviolet GALEX
\citep[Galaxy Evolution Explorer;][]{martinetal05-1,
  morrisseyetal05-1} and optical SDSS photometry to the fluxes predicted
from each solution (where the SDSS fluxes are derived directly from
the SDSS $u,g,r,i,z$ magnitudes). Spectroscopic fits that use 1D
atmosphere spectra models are known to systematically
overestimate surface gravities for cool ($\la$12,000\,K) white dwarfs
\citep{koesteretal09-1, tremblayetal11-1}. To overcome this effect we
applied the 3D corrections of \citet{tremblayetal13-1} to $\Teff$ and
$\log g$ determined above. Finally we obtained the masses of our white
dwarfs by interpolating the obtained $\Teff$ and $\log g$ values with
the tables of \cite{renedoetal10-1}.

\begin{figure}
\includegraphics[width=\columnwidth]{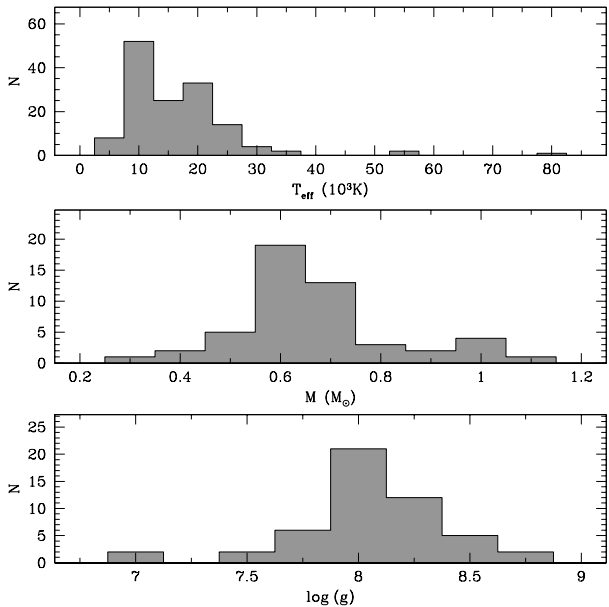}
\caption{\label{f-histo} From top to bottom: effective temperature, mass and surface gravity distributions of the new LAMOST DA white dwarfs identified in this work.}
\end{figure}

The $\Teff$, $\log g$ and masses we have determined are provided in
the online catalogue which accompanies this article (Table
\ref{Col_tab}). Inspection of the table reveals that these values are
subject to large uncertainties in many cases. This is due to the low
signal-to-noise ratio of many of the LAMOST
spectra. In Figure \ref{f-histo} we show the $\Teff$ distribution for DA white dwarfs with relative errors $\leq$10 per cent and the mass and $\log
g$ distributions for DA white dwarfs with errors in mass $< 0.075 M\odot$. This
results in 49 and 154 white dwarfs in the $\log
g$  and  $\Teff$ histograms respectively.
\newline \indent The mass distribution displays a broad and clear peak at 0.6-0.7\Msun\,, as
typically found in many previous
studies \citep[e.g.][]{liebertetal05-1, Kleinmanetal13-1,
  kepleretal15-1}. It also reveals the existence of both low-mass
($\la$0.5\Msun) white dwarfs that may harbour unseen companions
\citep{rebassa-mansergasetal11-1, kilicetal12-1} and high-mass
($\ga$0.8\Msun) white dwarfs, some of which may be the result of mergers \citep{giammicheleetal12-1,
  rebassa-mansergasetal15-1}.
\newline \indent The $\Teff$ distribution shows that our newly identified white dwarfs
have generally 10,000$-$20,000\,K, similar to the distribution obtained
from SDSS spectroscopically selected white dwarfs
\citep{kepleretal15-1}.

\subsection{Comparison with stellar parameters from SDSS spectra}
\begin{figure}
\includegraphics[width=\columnwidth]{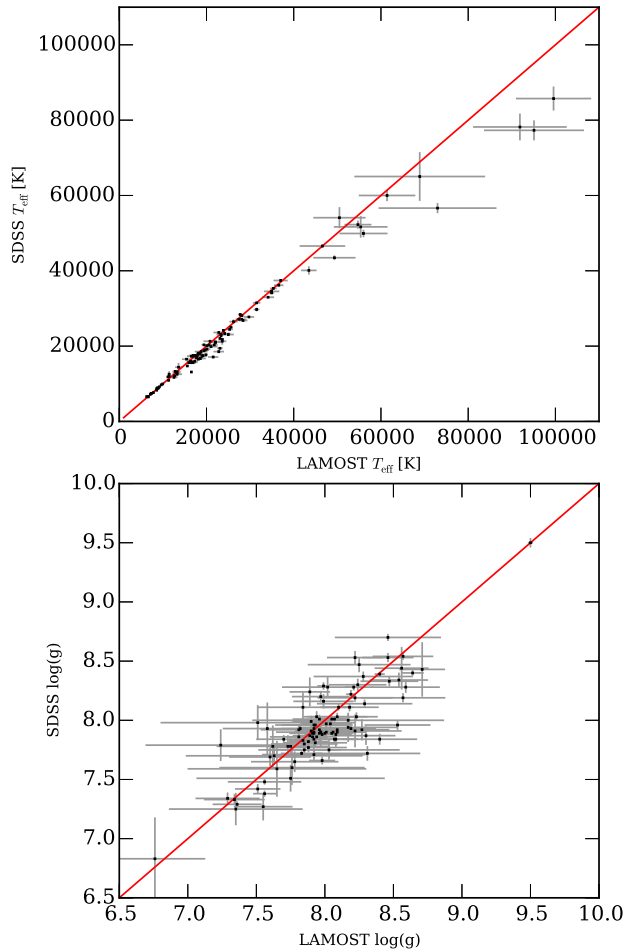}
\caption{\label{LAMOSTvsSDSS} Comparison of stellar parameters (top: $\Teff$, bottom: $\log g$) obtained by fitting the available SDSS and LAMOST spectra of 108 DA white dwarfs. Only objects which had a LAMOST spectrum with S/N $> 10$ were used. The red lines reflect a simple 1:1 relationship.}
\end{figure}

As mentioned in Section 3, 3500 objects from our initial sample have both SDSS and LAMOST spectra and 774 of them are white dwarfs (Table\,\ref{recap}). This sample of objects provides a useful opportunity to compare DA white dwarf the stellar parameter obtained from LAMOST spectra with those obtained from SDSS spectra. 
In this comparison we decided to limit ourselves to DA white dwarfs with LAMOST spectra with a signal-to-noise ratio $>10$. Applying these criteria results in a sample of 108 white dwarfs.
We derive  $\Teff$ and $\log g$ from both the SDSS spectra and the LAMOST spectra following the same procedure described in Section 6. 
Figure\,\ref{LAMOSTvsSDSS} shows that the uncertainties in the LAMOST stellar parameters are generally significantly larger than the SDSS ones, this is caused by systematically lower signal-to-noise ratio of the LAMOST spectra. 
In order quantify the discrepancy between LAMOST and SDSS $\Teff$ and $\log g$ we define a quantity $\tau$ :
\begin{equation}
\tau = \frac{SDSS\,\mathrm{value} - LAMOST\,\mathrm{value}}{\sqrt{SDSS\,\mathrm{\sigma^2}+LAMOST\,\mathrm{\sigma^2}}}
\end{equation}
We find that in 11 per cent of the cases the LAMOST $\Teff$ values are overestimated by more than 2 $\tau$ compared to the SDSS ones. 
When comparing the $\log g$ values only 5.6 per cent of the objects show a comparable disagreement ($\tau$ $>2$).
We can conclude that the stellar parameters computed using SDSS and LAMOST spectra are broadly in agreement.

\section{Summary and conclusions}

By cross matching all photometrically selected SDSS white dwarf
candidates from the \citet{gentilefusilloetal15-1} catalogue with the
over 4 million spectra currently provided by the LAMOST DR3 we
identified 309, bright ($g \leq19$) white dwarfs with available LAMOST spectra of which only 64 were previously published as LAMOST white dwarfs. We inspected and classified the remaining \Nwd\- objects
according to their spectral type and obtained \Teff, $\log g$ and masses
for the DAs (which make up 80 per cent of the sample).
We also find that \newwd\- of these white dwarfs were previously unknown and are therefore new discoveries. 
Since LAMOST targeted mainly object with  $g\simeq 14-16$, 
the sample of LAMOST white dwarfs presented here is limited in size.
However, because SDSS and
LAMOST follow different targeting strategies, the sample of LAMOST white
dwarfs is not affected by the known and heavy biases of
the SDSS white dwarf spectroscopic sample and constitute therefore a
valuable complementary addition.  

We also inspected the LAMOST available spectra of 864 additional SDSS
photometric sources from the \citet{gentilefusilloetal15-1} list and
classified them as non-white dwarfs, i.e. contaminant objects. We used
the spectroscopic samples of newly confirmed white dwarfs and
contaminants to test the reliability of the
\citet{gentilefusilloetal15-1} selection method. Even with the
relatively small size of our LAMOST spectroscopic sample we were able
to verify that the \citet{gentilefusilloetal15-1} \emph{probabilities
  of being a white dwarf} can be reliably used to select samples of
white dwarfs with completeness and efficiency close to 90 per cent.
These results show that similar searches to the one presented here may
be repeated in the near future (e.g with the forthcoming new data
release of LAMOST) in a much more efficient way by relying more on the
values of $P_\mathrm{WD}$ and therefore drastically reducing the
amount of data to inspect by eye. Since the
\citet{gentilefusilloetal15-1} selection method can be applied to any
multi-band photometric survey, future searches may also not be limited
to the SDSS footprint.

The \citet{gentilefusilloetal15-1} catalogue only includes SDSS white
dwarf candidates with $g \leq19$, $T_\mathrm {eff} \gtrsim 8000K$ and
reliable proper motions. Furthermore, LAMOST specifically targeted
areas of the sky (e.g the Galactic anti-center) which were instead
avoided by SDSS.  Hence, the catalogue of new LAMOST white dwarfs
presented here is not complete and should be considered as
complementary to the work done in \citet{gentilefusilloetal15-1}, as
well as a contribution to the total sample of known spectroscopic white dwarfs. 

In the near future further multi-band photometry of LAMOST targets will become available thanks to the Panoramic Survey Telescope \& Rapid Response System (Pan-STARRS, \citealt{panstarrs14-1}). The selection method for white dwarf candidates of \citet{gentilefusilloetal15-1} could be applied to the Pan-STARRS database. Such application will lead to a new search for LAMOST white dwarfs outside the SDSS footprint. 

\begin{table*}
\centering
\caption{\label{Col_tab} Format of the catalogue of  LAMOST white
  dwarfs. The full catalogue can be accessed online via the VizieR
  catalogue access tool.}
\begin{tabular}{lll}
\hline
\hline
Column No. & Heading & Description\\
\hline
1 & sdss name & SDSS objects name (SDSS + J2000 coordinates)\\
2 & ra & right ascension\\
3 & dec & declination\\
4 & probability & The \emph{probability of being a WD} computed for this object in \citet{gentilefusilloetal15-1}\\
5 & umag & SDSS $u$ band PSF magnitude\\
6 & umag err & SDSS $u$ band PSF magnitude uncertainty\\
7 & gmag & SDSS $g$ band PSF magnitude\\
8 & gmag err & SDSS $g$ band PSF magnitude uncertainty\\
9 & rmag & SDSS $r$ band PSF magnitude\\
10 & rmag err & SDSS $r$ band PSF magnitude uncertainty\\
11 & imag & SDSS $i$ band PSF magnitude\\
12 & imag err & SDSS $i$ band PSF magnitude uncertainty\\
13 & zmag & SDSS $z$ band PSF magnitude\\
14 & zmag err & SDSS $z$ band PSF magnitude uncertainty\\
15 & ppmra & proper motion in right ascension (mas/yr)\\
16 & ppmra err & proper motion in right ascension uncertainty (mas/yr)\\
17 & ppmdec & proper motion in right declination (mas/yr)\\
18 & ppmdec err & proper motion in right declination uncertainty (mas/yr)\\
19 & Simbad classification & Currently available Simbad classifications\\
20 & LAMOST spec ID & Unique spectra identifier composed of MJD (modified Julian date)\\
   &  & of the observation, plate ID, spectrograph ID and a fiber ID\\
21 & LAMOST class & classification of the object based on our visual inspection of its LAMOST spectra\\
22 & \Teff & effective temperature calculated for DA white dwarfs (sect. 6)\\
23 & \Teff\- err & uncertainty in the effective temperature calculated for DA white dwarfs (sect. 6)\\
24 & $\log g$ & surface gravity calculated for DA white dwarfs (sect. 6)\\
25 & $\log g$\- err & uncertainty in the surface gravity calculated for DA white dwarfs (sect. 6)\\
26 & M/\Msun & mass of the white dwarfs calculated for DA white dwarfs (sect. 6)\\
27 & M/\Msun\- err & uncertainty in mass the of the white dwarfs calculated for DA white dwarfs (sect. 6)\\
\hline
\end{tabular}

\end{table*}

\section*{Acknowledgements}

NPGF acknowledges the support of Science and Technology Facilities
Council (STFC) studentships.

ARM acknowledges financial support from the Postdoctoral Science
Foundation of China (grants 2013M530470 and 2014T70010) and from the
Research Fund for International Young Scientists by the National
Natural Science Foundation of China (grant 11350110496).

The research leading to these results has received funding from the
European Research Council under the European Union’s Seventh Framework
Programme (FP/2007-2013) / ERC Grant Agreement n. 320964 (WDTracer).

Guoshoujing Telescope (the Large Sky Area Multi-Object Fiber
Spectroscopic Telescope, LAMOST) is a National Major Scientific
Project which is built by the Chinese Academy of Sciences, funded by
the National Development and Reform Commission, and operated and
managed by the National Astronomical Observatories, Chinese Academy of
Sciences.

\textbf{We thank the anonymous referee for his quick and constructive review.}

Funding for SDSS-III has been provided by the Alfred P. Sloan
Foundation, the Participating Institutions, the National Science
Foundation, and the U.S. Department of Energy Office of Science. The
SDSS-III web site is http://www.sdss3.org/.

This research has made use of the SIMBAD database,
operated at CDS, Strasbourg, France.

The CSS survey is funded by the National Aeronautics and Space
Administration under Grant No. NNG05GF22G issued through the Science
Mission Directorate Near-Earth Objects Observations Program.  The CRTS
survey is supported by the U.S.~National Science Foundation under
grants AST-0909182 and AST-1313422.

\bibliographystyle{mn_new}


\end{document}